\documentclass[a4paper,11pt]{article}
\pdfoutput=1
\usepackage{jheppub}
\usepackage[utf8x]{inputenc}
\usepackage{graphicx}
\usepackage{float}
\usepackage[T1]{fontenc}
\usepackage{epsfig}
\usepackage{color}
\usepackage{amsmath,mathtools}
\usepackage{subfloat}
\usepackage{amsfonts}
\usepackage{braket}
\usepackage{cleveref}
\usepackage{epstopdf}
\usepackage{caption}
\usepackage{etoolbox}
\usepackage{orcidlink}
\usepackage{subcaption}
\usepackage{enumitem}
\usepackage{comment}
\usepackage[titletoc,toc,title]{appendix}
\usepackage{hyperref}
\hypersetup{
	colorlinks=true,
	linkcolor=blue,
	filecolor=red,      
	urlcolor=blue,
	citecolor=red
} 
\DeclareMathOperator{\sech}{sech}

\newcommand{\uc}{ _{_{\text{CFT}}} }

\title{\boldmath Entanglement, defects, and $T\bar{T}$ on a black hole background}

\author{Ankur Dey\,\orcidlink{0009-0001-1077-0442}}
\affiliation{Department of Physics, Indian Institute of Technology, Kanpur 208016, India}

\emailAdd{ankurd21@iitk.ac.in}

\abstract{In this article, we investigate the proposed duality between the island and the defect extremal surface (DES) prescriptions using the fine-grained entanglement entropy in Karch-Randall (KR) brane-world models with gravitating radiation baths. We consider the AdS$_3$ black string geometry and compute the entanglement entropy for radiation subsystems on an AdS$_2$ eternal black hole background using both the island and the DES prescriptions. We find an agreement between the two proposals for the island and the no-island phases, thus verifying the validity of the proposed duality. We further extend to a $T\bar{T}$ deformed AdS$_3$ black string geometry with a cut-off and find consistent results for both phases. We finally plot and compare the Page curves for the undeformed and deformed scenarios, and discuss the modifications due to $T\bar{T}$ deformation.}


\arxivnumber{2511.05256}

\begin{document} 
\maketitle
\flushbottom
\setcounter{tocdepth}{2}

\section{Introduction}\label{sec_intro}

The black hole information paradox has been an active research topic in semi-classical and quantum gravity for several decades. In recent developments, a possible resolution to this paradox was proposed which considered a toy model of a quantum field theory coupled to a semi-classical gravity, finally reproducing the infamous Page curve \cite{Page:1993wv,Page:1993df,Page:2013dx} for the Hawking radiation. The aptly named \textit{island formalism} involves the emergence of spacetime regions called 'islands' at late times in the entanglement wedge of the radiation bath, leading to the purification of the outgoing Hawking quanta \cite{Almheiri:2019hni,Almheiri:2019psf,Almheiri:2019qdq,Almheiri:2019psy,Almheiri:2019yqk,Almheiri:2020cfm,Geng:2025rov, Penington:2019npb}. Inspired by the quantum extremal surface (QES) prescription \cite{Engelhardt:2014gca}, the island formula was proposed for computing the fine-grained entanglement entropy of the Hawking radiation.

The double holographic prescription \cite{Almheiri:2019hni, Sully:2020pza, Rozali:2019day, Chen:2020uac, Chen:2020hmv, Grimaldi:2022suv, Suzuki:2022xwv, Geng:2020qvw, Geng:2020fxl, Geng:2021iyq, Geng:2021mic, Geng:2021hlu, Jain:2023xta, Geng:2024xpj} provides an interesting alternative perspective on the island formalism. It described a $d+1$-dimensional gravitational theory in a braneworld as a dual to a $d$-dimensional conformal field theory (CFT$_d$) coupled to semi-classical gravity on the End-of-the-world (EOW) brane. This coupled description, referred to as the brane or the lower-dimensional effective description, may be obtained via a partial reduction of the bulk Anti-de Sitter (AdS) geometry \cite{Deng:2020ent, Chu:2021gdb, Li:2021dmf, Basu:2022reu, Shao:2022wrm, Basu:2024xjq}. Furthermore, this gravity-matter system on the EOW brane is the holographic dual to a quantum mechanical system located at the boundary of the CFT$_d$, known as the boundary description. The calculation of entanglement entropy in this context aligns with the Ryu-Takayanagi (RT) prescription in the bulk AdS$_{d+1}$ geometry.

On a separate note, the author in \cite{Cardy:2004hm} introduced the boundary conformal field theory (BCFT) as a CFT defined on a domain with a boundary. The AdS/BCFT correspondence was subsequently described in \cite {Takayanagi:2011zk} and was further developed in recent years (see \cite{Fujita:2011fp, Sully:2020pza, Kastikainen:2021ybu,Izumi:2022opi,Cavalcanti:2018pta,Magan:2014dwa,Cavalcanti:2020rsp,Takayanagi:2020njm,Geng:2025efs}). The AdS$_3$/BCFT$_2$ correspondence describes an asymptotically AdS$_3$ geometry truncated by a codimension-1 EOW brane with Neumann boundary conditions as the holographic dual for a BCFT$_2$. Due to modified homology conditions of the RT surfaces in this framework, the holographic entanglement entropy formula involves extremal surfaces that terminate on the EOW brane \cite{Takayanagi:2011zk, Fujita:2011fp}.

This AdS$_3$/BCFT$_2$ scenario was further developed in \cite{Deng:2020ent}, where the authors considered the bulk as a defect spacetime with conformal matter on the EOW brane. In this case, the gravity region relevant to determining the QES is localized only on the EOW brane and emerges via a partial Randall-Sundrum reduction of the bulk geometry. In the lower-dimensional effective description, transparent boundary conditions were imposed at the interface of the gravitational and non-gravitational regions of the CFT$_2$. This description, known as the defect extremal surface (DES) prescription, was proposed as a holographic counterpart of the island formula. Computations of the fine-grained entanglement entropy using the DES formula and the boundary QES prescription were shown to agree. Subsequently, the authors in \cite{Chu:2021gdb} employed the DES formula to compute the entanglement entropy for subsystems in radiation baths associated with a $2d$ eternal black hole.

To assess the robustness of the island framework and its double holographic generalizations, it is crucial to examine them in a less conventional gravitational setting. The solvable nature of CFTs deformed by an irrelevant operator formed from the stress-energy tensors \cite{Zamolodchikov:2004ce, Cavaglia:2016oda, Smirnov:2016lqw} creates a suitable environment for this purpose. These irrelevant deformations, termed as $T\bar{T}$ deformations and extensively studied in the literature \cite{Shyam:2017znq, Kraus:2018xrn, Cottrell:2018skz, Taylor:2018xcy, Hartman:2018tkw, Shyam:2018sro, Caputa:2019pam, Giveon:2017myj, Asrat:2017tzd, Donnelly:2018bef, Lewkowycz:2019xse, Chen:2018eqk, Banerjee:2019ewu, Jeong:2019ylz, Murdia:2019fax, Park:2018snf, Asrat:2019end, He:2019vzf, Grieninger:2019zts, Khoeini-Moghaddam:2020ymm, Basu:2024bal, Basu:2023aqz, Basu:2023bov, Basu:2024enr,Grieninger:2023knz,Chang:2024voo,Basu:2024xjq,Basu:2025exh,Basu:2025fsf,Hirano:2025alr,Hirano:2025tkq,He:2025ppz,Pant:2024eno}, introduce a finite cut-off in the AdS$_3$ geometries \cite{McGough:2016lol} with Dirichlet boundary conditions.\footnote{Note that alternate holographic proposals for $T\bar{T}$ deformed theories, such as \cite{Hirano:2020nwq, Kraus:2018xrn, Guica:2019nzm,Dubovsky:2018bmo, Tolley:2019nmm,Hirano:2025cjg}, exist in the literature and are expected to coincide under certain limits.}

In \cite{Deng:2023pjs,Deng:2024dct}, the authors computed the entanglement entropy for bipartite pure states using both the island and the DES formula within a cut-off AdS$_3$ spacetime truncated by an EOW brane. This setup involves two boundaries- an EOW brane characterizing the AdS/BCFT correspondence, and a finite cut-off surface that arises due to $T\bar{T}$ deformed theories. These surfaces differ fundamentally due to the distinct boundary conditions imposed and introduce subtleties in the standard AdS/BCFT correspondence, which require careful analysis. In the boundary description, this construction is equivalent to a $T\bar{T}$ deformed BCFT$_2$ obtained via partial dimensional reduction of the braneworld gravity glued to a non-gravitating CFT with $T\bar{T}$ deformation \cite{Deng:2023pjs}.

In \cite{Deng:2020ent}, the proposed duality between the island formula and the DES prescription has been verified using the entanglement entropy for simple braneworld models where the induced geometry on the EOW brane was Poincaré AdS$_2$ and the BCFT was on a flat and non-gravitating background. Further verification of this duality using mixed state entanglement measures like the entanglement negativity and the reflected entropy was subsequently carried out in \cite{Basu:2022reu, Shao:2022wrm, Li:2021dmf}, and for $T\bar{T}$ deformed scenarios in \cite{Deng:2023pjs,Deng:2024dct, Basu:2024xjq}. However, it is crucial to verify this duality in more general settings where the induced metric on the brane and the BCFT background might not be so simple. In \cite{Wang:2021xih} the authors considered a gravitational FRW cosmology on the EOW brane coupled to a flat CFT on the asymptotic boundary, and verified the equivalence between the island and the DES prescriptions using the entanglement entropy. 

Although significant progress has been made, a systematic test for this duality in settings where both the EOW brane and the radiation bath reside on non-trivial geometries, as well as in the presence of irrelevant deformations such as the $T\bar{T}$ deformation, is still an open problem. In this article, we address this interesting issue and investigate the proposed duality between the island and the DES prescription for braneworld models where the radiation bath and the EOW brane reside on a black hole geometry. In particular, we consider the model described in \cite{Geng:2021mic,Geng:2022dua}, where the authors consider a bulk AdS$_3$ black string geometry truncated by a Karch-Randall (KR) brane, dual to a BCFT$_2$ considered on an eternal AdS$_2$ black hole background. In this model, the metric induced on the brane is also an eternal AdS$_2$ black hole. We modify this model to include conformal matter localized on the EOW brane, and utilizing the island and the DES prescription, we compute the entanglement entropy for subsystems considered in radiation baths associated with an eternal AdS$_2$ black hole. We show that the results obtained using the DES prescription align with those obtained from the island formula, and provide a strong verification for the proposed duality. We further extend our analysis to an AdS$_3$ black string geometry with a finite cut-off arising from the $T\bar{T}$ deformation and truncated by the EOW brane, and compute the entanglement entropy up to first order corrections for subsystems in the radiation baths. Once again, we demonstrate an agreement between the results obtained using the bulk DES and the island prescriptions.

The remainder of the article is as follows. In \cref{sec_review} we briefly discuss some necessary concepts, such as the Defect AdS$_3$/BCFT$_2$, the island and the DES formula for the entanglement entropy, the braneworld model of a holographic BCFT$_2$ located on an AdS$_2$ black hole background, and $T\bar{T}$ deformation. In \cref{sec_ee}, we discuss the computations of the entanglement entropy for subsystems in the radiation baths in this braneworld model, utilizing both the island and the DES prescription. Subsequently, we determine the correction to the entanglement entropy in the cut-off AdS$_3$ black string model arising from $T\bar{T}$ deformation truncated by the EOW brane in \cref{sec_ett}. We generate and compare Page curves for both the undeformed and the $T\bar{T}$ deformed scenarios in \cref{sec_pc}, and finally, we conclude by summarizing our results and discussing their implications in \cref{sec_summary}.

\section{Review of earlier literature}\label{sec_review}

\subsection{Defect AdS$_3$/BCFT$_2$}\label{ssec_defect}

\begin{figure}[t]
\centering
\includegraphics[scale=1.2]{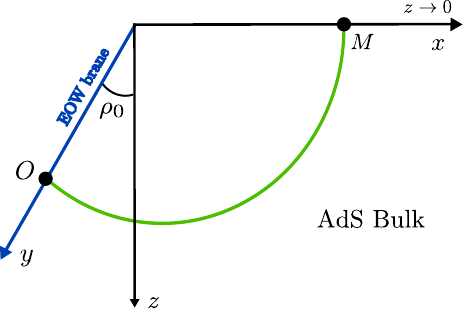}
\caption{Holographic dual of a BCFT$_2$ defined on a half plane $(x > 0)$. The EOW brane is considered to be at a constant hyperbolic angle $\rho_0$. Figure modified from \cite{Deng:2020ent}.}\label{fig_bcft}
\end{figure}

In this section we briefly review the AdS$_3$/BCFT$_2$ correspondence first proposed in \cite{Takayanagi:2011zk,Fujita:2011fp}. A boundary conformal field theory (BCFT) is defined as a conformal field theory (CFT) with a boundary, where conformal boundary conditions are applied. The holographic dual of a BCFT \cite{Cardy:2004hm} is an asymptotically AdS spacetime truncated by a co-dimension 1 end-of-the-world (EOW) brane where Neumann boundary conditions are applicable (see \cref{fig_bcft}). The bulk action in gives as \cite{Takayanagi:2011zk,Fujita:2011fp}
\begin{align}
I_0 = - \frac{1}{16 \pi G_N} \int_N \sqrt{g} (R-2 \Lambda) -\frac{1}{8 \pi G_N} \int_Q \sqrt{h} (K-(d-1)T),
\end{align}
where $N$ and $g$ represent the bulk AdS spacetime and its metric, while $Q$ and $h$ represent the EOW brane and the metric induced on the brane. $K$ is the extrinsic scalar curvature of the EOW brane and $T$ is the tension of the brane. By variation of the above action with respect to the induced metric $h_{ab}$ and implementation of the Neumann boundary condition, we arrive at the expression
\begin{align}
K_{ab} = (K-T)h_{ab}.
\end{align}
The AdS$_3$ spacetime may be expressed in terms of AdS$_2$ foliations as follows
\begin{align}\label{met_fol}
ds^2 = d \rho^2 + \cosh ^2 \rho \left( \frac{d \tau^2 + dy^2}{y^2} \right),
\end{align}
where $y$ is the radial coordinate alone the foliation and $\rho$ is the hyperbolic angle between the foliation and the normal to the asymptotic boundary. Note that we have assumed the AdS length scale to be unity for simplicity. Assuming that the EOW brane is located at some constant $\rho = \rho_0$, the metric induced on this brane may be expressed as a conformally flat metric with a conformal factor $\Omega_y$ as
\begin{align}\label{met_brane}
ds^2 _{brane} = \Omega^2_y (d \tau^2 + dy^2) = \Omega^2_y \,\ ds^2 _{flat}, \qquad \Omega_y = \bigg| \frac{1}{y \sech \rho_0 } \bigg|.
\end{align}
It is also fairly simple to show that
\begin{align}
K_{ab} = \tanh \rho_0 \,\ h_{ab}, \qquad T = \tanh \rho_0.
\end{align}

This AdS$_3$/BCFT$_2$ setup was further developed in \cite{Deng:2020ent,Chu:2021gdb} where the authors incorporated conformal matter localised on a tensionless EOW brane. This induces a tension on the brane, which result in the brane to relocate to some constant angle dependent on the tension. The bulk action in this modified setup is given by
\begin{align}
I = I_0 + \int_{Q} \sqrt{h} \mathcal{L}_Q,
\end{align}
where $\mathcal{L}_Q$ is the Lagrangian corresponding to the conformal matter on the brane.\footnote{See \cite{Chen:2020uac} for some examples.} Once again by variation of this modified action with respect to the induced metric $h_{ab}$ and implementation of the Neumann boundary condition we arrive at \cite{Takayanagi:2011zk,Deng:2020ent,Chu:2021gdb}
\begin{align}
K_{ab} = (K-T)h_{ab}+8 \pi G_N \langle T_{ab} \rangle, \qquad \langle T_{ab} \rangle=\frac{2}{\sqrt{h}} \frac{\partial \mathcal{L} _Q}{\partial h_{ab}}.
\end{align}
Here $\langle T_{ab} \rangle$ is the expectation value of the stress energy tensor of the conformal matter on the brane, and is proportional to the induced metric $h_{ab}$ in this case. These conformal degrees of freedom on this brane do not backreact on the bulk geometry, but rather modify the brane's intrinsic properties. As a result, the EOW brane may be treated as a non-dynamical lower-dimensional defect in the bulk AdS$_3$ spacetime.

\subsection{Entanglement Entropy and the DES formula}\label{ssec_eedes}

In the modified scenario of defect AdS$_3$/BCFT$_2$ discussed in the previous subsection, the entanglement entropy of a subsystem $A$ undergoes modifications to incorporate contributions from the conformal matter localised on the brane. In this framework the fine grained entanglement entropy of a subsystem $A$ defined in the radiation bath is computed using the island formula described in \cite{Almheiri:2019hni,Penington:2019npb, Almheiri:2019psf} as
\begin{align}\label{eq_eeis}
S_{Is} = \text{min} \,\ \text{ext} _X \Big\{ S_{eff}(A \cup I_S(A)) + S_{area}(X) \Big\}, \qquad X=\partial I_S(A)
\end{align}
where $I_S(S)$ represents the entanglement entropy island on the EOW brane corresponding to the subsystem $A$, while $X$ denotes the boundary of the island $I_S(S)$.

From the holographic perspective, the Defect Extremal Surface (DES) formula was introduced as a holographic counterpart of the island formula in \cite{Deng:2020ent}, where the authors enhanced the Quantum Extremal Surface (QES) formula to incorporate contributions from the bulk defect matter localized on the EOW brane.  This leads to the modification of the Ryu-Takayanagi (RT) formula for the holographic entanglement entropy to the DES formula as
\begin{align}\label{eq_eedes}
S_{DES} = \text{min} \,\ \text{ext} _{\Gamma_A , X} \Big\{ S_{RT}(\Gamma_A) + S_{defect}(D) \Big\}, \qquad X=\Gamma_A \cap D,
\end{align}
where $\Gamma_A$ is the RT surface and $D$ is the defect region along the EOW brane. The term $S_{defect}(D)$ is the contribution to the entanglement which arises due to the conformal matter localized on the brane.

\subsection{Holographic BCFT$_2$ in a Black Hole background}\label{ssec_bcftbh}

In \cite{Geng:2021mic,Geng:2022dua}, the authors examined the entanglement entropy of subsystems considered on the asymptotic boundary of an AdS$_3$ black string geometry truncated by a Karch-Randall (KR) brane. The bulk geometry is described by the metric
\begin{align}
ds^2=\frac{\cosh ^2 \rho}{u^2} \bigg[ - \left( 1-\frac{u}{u_h} \right) dt^2 + \frac{du^2}{\left( 1-\frac{u}{u_h} \right)} \bigg] +d \rho ^2,
\end{align}
where $\rho$ is a hyperbolic coordinate such that $\rho \in [- \infty,\infty]$. The asymptotic boundary is defined at $\rho = - \infty \cup \infty$, and the KR brane is assumed to be embedded on a constant $\rho = -\rho_B$ slice. The accessible region extends from $\rho = -\rho_B$  to $\rho = \infty$. The geometry on each constant $\rho$ slice is an eternal AdS$_2$ black hole. The dual BCFT$_2$ is thus living on an eternal AdS$_2$ black hole background with two asymptotic boundaries, with conformal boundary conditions applicable at $u=0$. Utilizing the relations
\begin{align}\label{eq_embed}
& X_0=\frac{2 u_h-u}{u} \cosh \rho, 
\qquad \qquad \qquad \qquad X_1=\frac{2 \sqrt{u_h^2-uu_h}}{u} \sinh \frac{2 \pi t}{\beta} \cosh \rho, \notag \\
& X_2=\frac{2 \sqrt{u_h^2-uu_h}}{u} \cosh \frac{2 \pi t}{\beta} \cosh \rho,
\qquad X_3=\sinh \rho,
\end{align}
the bulk geometry may be embedded in a higher $\mathbb{R}^{2,2}$ geometry
\begin{align}
ds^2 = \eta _{AB} dX^A dX^B, \qquad \eta _{AB} =\text{diag}(-1,-1,1,1),
\end{align}
with an additional constraint of
\begin{align}
X_A X^A =-1.
\end{align}
Here $\beta=4 \pi u_h$ is the inverse Hawking temperature. Using the above embedding relations, the holographic entanglement entropy for radiation subsystems considered on both the left and right copies of the TFD state corresponding to an eternal black hole was evaluated in \cite{Geng:2022dua} as
\begin{align}\label{eq_eecft}
S_A = 
\begin{cases}
\frac{c}{3} (\rho_{\epsilon} + \rho_b) 
& \text{Boundary channel,} \\
\frac{c}{6}\left( \log \left[ \frac{u_h}{L R} \left( \Delta_L + \Delta_R + 2\sqrt{\Delta_L \Delta_R} \cosh \frac{t}{u_h}\right)\right] +2 \rho_{\epsilon} \right) \qquad
& \text{Bulk channel,}
\end{cases}
\end{align}
where $\Delta_L=u_h-L$, $\Delta_R=u_h-R$, and $\rho_{\epsilon}$ is related to the UV cut-off $\epsilon$ as $\rho_{\epsilon}=\log \frac{2}{\epsilon}$. These results were also reproduced from the boundary description using BCFT correlators of twist fields in a black hole background in \cite{Geng:2022dua}.

% \textbf{Double holographic perspective: } The boundary description discussed above defines a BCFT$_2$ in a thermofield double (TFD) on an AdS$_2$ black hole background with boundary degrees of freedom localized at $u=0$. An equivalent bulk description is the dual AdS$_3$ black string geometry truncated by a KR brane \cite{Takayanagi:2011zk}. The brane angle $-\rho_B$ specifies the boundary degrees of freedom. Another equivalent standpoint, called the effective lower-dimensional description, arises in the context of braneworld holography \cite{Karch:2000ct, Karch:2000gx} by integrating the bulk degrees of freedom along the $\rho$ direction. In this perspective, the CFT$_2$ is coupled to a hybrid manifold that consists of two AdS$_2$ black holes interacting through a shared interface, where one acts as a radiation reservoir and collects the Hawking radiation emitted by the other.  *** \textbf{It is needed?} ***

%
%
%
%
%

\subsection{$T\bar{T}$ deformation}\label{ssec_ttbar}

We now briefly review $T\bar{T}$ deformation in CFTs and its holographic dual. The $T\bar{T}$ deformation of a $2d$ conformal field theory considered on a flat space is defined as \cite{Cavaglia:2016oda,Smirnov:2016lqw}
\begin{align}
\frac{d I_{QFT}^{(\mu)}}{d \mu}=-2 \pi \int d^2x (T\bar{T})_{\mu}, \qquad I_{QFT}^{(\mu)}\Bigg|_{\mu=0}=I_{CFT},
\end{align}
where $I$ represents the Lorentzian action and $\mu$ is the deformation parameter. The $T\bar{T}$ deformation operator $(T\bar{T})$ may be defined in terms of the components of the stress energy tensor $T^{ab}$ as
\begin{align}
(T\bar{T})=\frac{1}{8} \left( T_{ab}T^{ab}-(T^a_a)^2 \right).
\end{align}
This operator is shown to satisfy the factorization formula as \cite{Zamolodchikov:2004ce}
\begin{align}
\langle T\bar{T} \rangle = \frac{1}{8} \left( \langle T_{ab} \rangle \langle T^{ab} \rangle - \langle T_a^a \rangle ^2 \right).
\end{align}
The corresponding bulk dual of the $T\bar{T}$ deformed CFT$_2$ in a flat space was proposed by the authors in \cite{McGough:2016lol,Lewkowycz:2019xse} to be a quantum gravity in AdS$_3$ with a radial cut-off as 
\begin{align}
ds^2 = \frac{1}{z^2} \left( -dt^2 + dx^2 + dz^2 \right), \qquad z > z_c,
\end{align}
where Dirichlet boundary conditions are imposed at $z=z_c$. The deformation parameter is related to this radial cut-off $z_c$ as\cite{Lewkowycz:2019xse}
\begin{align}\label{eq_defpar}
\mu = 8 G_N z_c^2 = \frac{12}{c} z_c^2.
\end{align}

\section{Entanglement entropy using the DES prescription}\label{sec_ee}

In this section we consider conformal matter located on the EOW brane and investigate the entanglement entropy in the effective lower-dimensional description discussed in the previous section for a radiation subsystem in the AdS$_3$ black string model. We also show that the results are consistent with those obtained using the DES formula in the bulk description.

\subsection{Including brane defects}

Recall the metric of an AdS$_3$ black string geometry
\begin{align}\label{met_bs}
ds^2= d \rho ^2 + \frac{\cosh ^2 \rho}{u^2} \bigg[  \left( 1-\frac{u}{u_h} \right) d\tau_{bs}^2 + \frac{du^2}{\left( 1-\frac{u}{u_h} \right)} \bigg] ,
\end{align}
which is now expressed in terms of the Euclidean time coordinate $\tau_{bs}$.\footnote{The subscript '$bs$' denotes 'black string', making $\tau_{bs}$ the Euclidean time coordinate in the black string coordinates. Subsequently we will denote the Euclidean time coordinate in the Poincar\'e coordinates as $\tau_p$.} Using the conformal transformation
\begin{align}\label{trans_1}
\omega =\frac{u_h e^{\frac{i \tau_{bs} }{2 u_h}}}{\sqrt{1-\frac{u}{u_h}}},
\end{align}
we obtain the thermofield double state on the metric
\begin{align}\label{eq_cf1}
ds^2= d \rho ^2 + \frac{4 u_h^2 \cosh^2 \rho }{(u_h^2-\omega \bar{\omega})^2} d\omega d\bar{\omega}.
\end{align}
The conformal boundary is mapped to the circle $\omega \bar{\omega}=u_h^2$ in these new coordinates. Utilizing the conformal transformation
\begin{align}\label{trans_2}
\omega= \frac{u_h}{\tau_p +i y -\frac{i}{2}}-i u_h,
\end{align}
we may further map the geometry to the Euclidean AdS$_3$/BCFT$_2$, where the background metric and the conformal factor now have the form
\begin{align}\label{eq_cf2}
ds^2 = d \rho ^2 + \Omega_y^2 (d \tau_p^2 + dy^2), \qquad \Omega_y= \Big| \frac{\cosh \rho}{y} \Big|,
\end{align}
which is the same as \cref{met_fol}. The conformal boundary is now located at $y=0$. Finally using the transformations
\begin{align}\label{trans_3}
z = y \sech \rho , \qquad x = y \tanh \rho , 
\end{align}
we obtain the well know Poincar\'e metric
\begin{align}
ds^2 = \frac{d \tau_p^2 + dx^2 + dz^2}{z^2}.
\end{align}

\subsection{Entanglement Entropy}\label{ssec_ee}

We now compute the entanglement entropy for a subsystem $A \equiv [L,u_h] \cup [u_h,R]$ on a constant time slice in the radiation baths of the eternal black hole model described previously. The endpoints of $A$ are denoted by the points $P \equiv (t,L,\rho_{\epsilon})$ and $Q \equiv (-t+i 2 \pi u_h,R,\rho_{\epsilon})$ on the asymptotic boundary.\footnote{Note that to account for the reversal of the timelike Killing vector as one crosses black string horizon from the left to the right partition, we take the time coordinate on the right partition $t_R$ to be related to that of the left partition $t_L$ as $t_R \to -t_L + i 2 \pi u_h$.} For simplicity of computation we begin in the Poincar\'e coordinates, where the corresponding endpoints are now at $P \equiv (\tau_1,x_1,z_{\epsilon_1})$ and $Q \equiv (\tau_2,x_2,z_{\epsilon_2})$ such that $z_{\epsilon_1}, z_{\epsilon_2} \to 0$. We later transform the final result using \cref{trans_1,trans_2,trans_3} to express the entanglement entropy in the AdS black string coordinates once again. Depending on the distance of subsystem $A$ from the EOW brane considered at a constant $\rho = - \rho_b$, we come across two distinct entanglement entropy phases, which are discussed below.

\subsubsection{Island Phase}

\begin{figure}[t]
\centering
\includegraphics[scale=0.8]{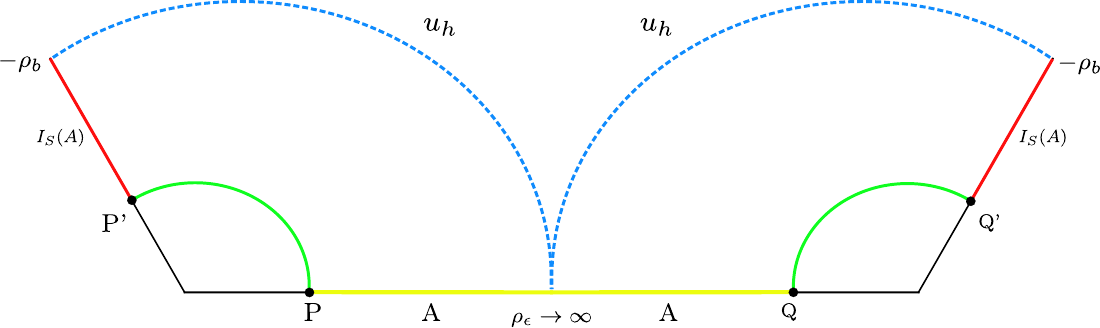}
\caption{The Island phase of the entanglement entropy. The radiation subsystem $A$ is denoted by the yellow subregions on the asymptotic boundary, while the corresponding entanglement entropy island $I_S(A)$ are illustrated by the red regions on the EOW brane. The green curves represent the RT surfaces.}\label{fig_eei}
\end{figure}

\textbf{Field theory computation :} In this phase the points $P$ and $Q$ are assumed to be close to the brane such that the subsystem $A$ on the radiation bath has a corresponding island region on the brane, as illustrated in \cref{fig_eei}. Under these circumstances the entanglement entropy may be determined using the island formula in \cref{eq_eeis}, where the effective entanglement entropy may be expressed in terms of the 4-point correlators of twist fields at the points $P,P',Q$ and $Q'$, which in the large $c$ limit factorizes as 
\begin{align}
\langle \sigma_n (P) \bar{\sigma}_n (P') \sigma_n (Q') \bar{\sigma}_n (Q) \rangle \uc \approx \langle \sigma_n (P) \bar{\sigma}_n (P') \rangle \uc \langle \sigma_n (Q') \bar{\sigma}_n (Q) \rangle \uc.
\end{align}
This allows us to express the first term in \cref{eq_eeis} as
\begin{align}
S_{eff} = \lim_{n \to 1} \frac{1}{1-n} \Big[ \log \Omega_{P'}^{-2 h_n}\langle \sigma_n (P) \bar{\sigma}_n (P') \rangle \uc + \log \Omega_{Q'}^{-2 h_n}\langle \sigma_n (Q') \bar{\sigma}_n (Q) \rangle \uc \Big],
\end{align}
where $\Omega_{P'}$ and $\Omega_{Q'}$ are the conformal factor at the points $P'$ and $Q'$ respectively, which are determined using \cref{eq_cf2}. Assuming $P' \equiv (-y_{_{P'}},\tau_{_{P'}})$ and $Q' \equiv (-y_{_{Q'}},\tau_{_{Q'}})$, with the coordinate $y$ running along the brane, the effective entanglement entropy may be obtained using the standard form of the 2-point correlator as
\begin{align}
S_{eff} & = \frac{c}{6} \left( \log \left[ \frac{(y_{_{P'}}+x_1)^2+(\tau_1-\tau_{_{P'}})^2}{\epsilon _y z_{\epsilon_1} }\right] + \log \left[ \frac{(y_{_{Q'}}+x_2)^2+(\tau_2-\tau_{_{Q'}})^2}{\epsilon _y z_{\epsilon_2} }\right] \right. \notag \\
& \qquad \left. + \log  \frac{1}{y_{_{P'}} \sech \rho _b} + \log  \frac{1}{y_{_{Q'}} \sech \rho _b} \right)
\end{align}
where $\epsilon_y$ denotes the UV cut-off on the brane. Adding to it the area term provided in \cite{Deng:2020ent}, the generalized entanglement entropy in this scenario may then be expressed as 
\begin{align}
S_{gen} = S_{eff} + 2 \times \frac{1}{4 G_N} \rho_b ,
\end{align}
which is extremized at $y_{_{P'}}=x_1, \tau_{_{P'}}=\tau_1$ and $y_{_{Q'}}=x_2, \tau_{_{Q'}}=\tau_2$. Substituting this back into the expression of $S_{gen}$ we finally obtain the entanglement entropy for the island phase as
\begin{align}\label{eq_eeip}
S(A) & = \frac{c}{6} \left( \log \frac{2 x_1}{z_{\epsilon_1}} + \log \frac{2 x_2}{z_{\epsilon_2}} + 2 \rho_b + 2 \log  \frac{2}{\epsilon_y \sech \rho _b} \right) \notag \\
& = \frac{c}{3} \left( \rho_{\epsilon} + \rho_b + \log  \frac{2}{\epsilon_y \sech \rho _b}  \right)
\end{align}
where in the second step we utilize the transformations in \cref{trans_1,trans_2,trans_3} to express the entanglement entropy in the AdS black string coordinates. The results agrees with the entanglement entropy in the boundary channel in \cref{eq_eecft}, with an additional term arising due to the contributions from the conformal matter on the EOW brane.

\vspace{0.5cm}

\textbf{Bulk computation :} From the bulk perspective the holographic entanglement entropy can be computed using the DES formula \cref{eq_eedes}, where the first term can be expressed as the combined lengths of the two RT surfaces $PP'$ and $QQ'$ denoted by the green curves in \cref{fig_eei}. The geodesic length between two arbitrary points $Y_i$ and $Y_j$ in AdS$_3$ may be expressed in the Poincar\'e coordinates as
\begin{align}\label{eq_geol}
L_{ij}=\cosh^{-1} \left[ - \xi_{ij} \right],
\end{align}
where $\xi_{ij}$ may be expressed in terms of the inner products of the position vectors of $Y_i$ and $Y_j$ as
\begin{align}
\xi_{ij} = Y_i \cdot Y_j = - \frac{(x_i-x_j)^2 + (\tau_i - \tau_j)^2 + z_i^2 + z_j^2}{2 z_i z_j}.
\end{align}
Using \cref{eq_geol} the combined length of $PP'$ and $QQ'$ may be determined as
\begin{align}
S_{RT} = & \frac{1}{4 G}  \cosh ^{-1}\left[ \frac{2 x_1 y_{_{P'}} \tanh \rho _b+x_1^2+y_{_{P'}}^2+(\tau_1-\tau_{_{P'}})^2+z_{\epsilon_1}^2}{2 z_{\epsilon_1} y_{_{P'}} \sech \rho _b}\right] \notag \\
& + \frac{1}{4 G}  \cosh ^{-1}\left[ \frac{2 x_2 y_{_{Q'}} \tanh \rho _b+x_2^2+y_{_{Q'}}^2+(\tau_2-\tau_{_{Q'}})^2+z_{\epsilon_2}^2}{2 z_{\epsilon_2} y_{_{Q'}} \sech \rho _b}\right],
\end{align}
where using \cref{trans_3} the coordinates of the bulk points $P'$ and $Q'$ are expressed as $P' \equiv (\tau_{_{P'}},-y_{_{P'}} \tanh \rho_b, y_{_{P'}} \sech \rho_b)$ and $Q' \equiv (\tau_{_{Q'}},-y_{_{Q'}} \tanh \rho_b, y_{_{Q'}} \sech \rho_b)$. The contribution from the defect term in \cref{eq_eedes} may be obtained directly from \cite{Deng:2020ent}, and the generalized entanglement entropy may then be described as
\begin{align}
S_{gen} = S_{RT}+\frac{c}{3} \log  \frac{2}{\epsilon_y \sech \rho _b}.
\end{align}
Upon extremization of the above expression we obtain $y_{_{P'}}=x_1, \tau_{_{P'}}=\tau_1$ and $y_{_{Q'}}=x_2, \tau_{_{Q'}}=\tau_2$, which when put back into $S_{gen}$ gives the final holographic entanglement entropy in the island phase (upon utilizing the Brown-Henneaux relation $c=\frac{3}{2 G}$) as \cref{eq_eeip}.

\subsubsection{No-Island Phase}\label{ssec_eeni}

\begin{figure}[t]
\centering
\includegraphics[scale=0.8]{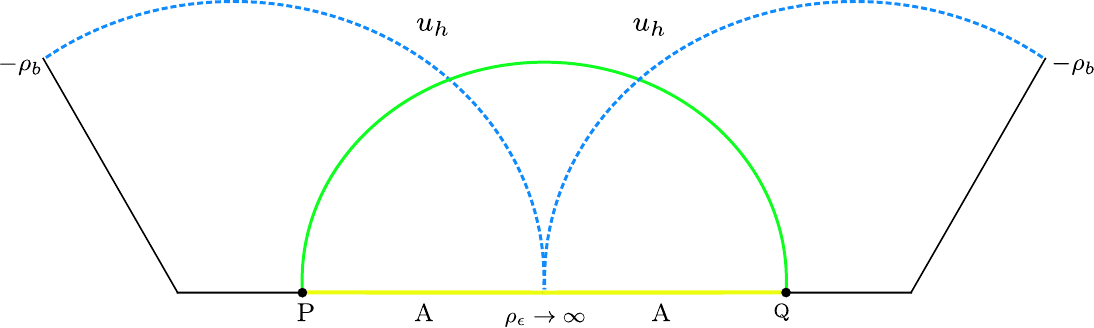}
\caption{The No-Island phase of the entanglement entropy. The radiation subsystem is once again denoted by the yellow subregions on the asymptotic boundary. The RT surface in this scenario is a Hartman-Maldacena type surface, and is represented by the green curve.}\label{fig_eeni}
\end{figure}

\textbf{Field theory computation :} In this phase the points $P$ and $Q$ are considered to be far away from the brane such that the subsystem $A$ on the radiation bath has no corresponding island region on the brane. This is illustrated in \cref{fig_eeni}. The entanglement entropy in this scenario is once again obtained using \cref{eq_eeis}, where the effective entanglement entropy is determined in terms of the 2-point correlator of twist fields at the points $P$ and $Q$ as
\begin{align}\label{eq_pqcorr}
S_{eff} = \lim_{n \to 1} \frac{1}{1-n} \log \langle \sigma_n (P) \bar{\sigma}_n (Q) \rangle \uc.
\end{align}
Due to the absence of islands regions on the brane corresponding to the radiation subsystem $A$, the area term in \cref{eq_eeis} does not contribute in this case. As a result of which the generalized entanglement entropy is simply the effective term as $S_{gen}=S_{eff}$. The entanglement entropy is then obtained as
\begin{align}
S(A) = \frac{c}{6} \log \left[ \frac{(x_2-x_1)^2+(\tau_2-\tau_1)^2}{z_{\epsilon_1} z_{\epsilon_2}} \right],
\end{align}
where once again using the transformations described in \cref{trans_1,trans_2,trans_3} may be utilized to express the result in the AdS black string coordinates as
\begin{align}\label{eq_eenip}
S(A) = \frac{c}{6} \log \left[ \frac{u_h}{L R} \left( \Delta_L+\Delta_R + 2 \sqrt{\Delta_L \Delta_R} \cosh \frac{t}{u_h} \right) \right]+ \frac{c}{3} \rho_\epsilon.
\end{align}
We observe that this result matches exactly with the bulk channel results in \cref{eq_eecft}. This is to be expected, since the conformal matter do not contribute to the entanglement entropy due to the absence of any corresponding entanglement entropy island on the brane.

\vspace{0.5cm}

\textbf{Bulk computation :} The holographic entanglement entropy in this case is once again computed using the DES formula given in \cref{eq_eedes}, where the first term is expressed in terms of the geodesic length of the Hartman-Maldacena (HM) surface $PQ$ represented by the green curve in \cref{fig_eeni}, which is given as
\begin{align}
S_{RT}=\frac{1}{4 G} \cosh ^{-1} \left[ \frac{(x_2-x_1)^2+(\tau_2-\tau_1)^2+z_{\epsilon_1}^2+z_{\epsilon_2}^2}{2 z_{\epsilon_1} z_{\epsilon_2}} \right].
\end{align}
As elaborated earlier, due to the absence of corresponding island regions on the EOW brane, second term in \cref{eq_eedes} does not contribute in this case. As a result we have $S_{gen}=S_{RT}$, and the final holographic entanglement entropy is then obtained as \cref{eq_eenip} on application of the Brown-Henneaux relation.

\section{Correction to entanglement entropy due to $T\bar{T}$ deformation}\label{sec_ett}

Having verified that the duality between the island prescription and the bulk DES formula hold for braneworld models where the radiation baths are on a black hole background, in this section we now extend our analysis to a $T\bar{T}$ deformed CFT$_2$ at the boundary, and compute the linear order correction to the entanglement entropy of a radiation subsystem in the AdS$_3$ black string model.

\subsection{Including $T\bar{T}$ deformation}

The tensionless brane in the bulk corresponds to a fixed point at $x=0$ in the asymptotic boundary. By ensuring that the deformed theory on the asymptotic boundary preserves the $Z_2$ symmetry at this fixed point, we may define a $T\bar{T}$ deformed theory with a boundary. Though the holographic dual of a $T\bar{T}$ deformed CFT$_2$ in a flat space is still AdS$_3$ with a radial cut-off, the introduction of a $Z_2$ quotient modifies the bulk metric such that 
\begin{align}\label{met_ttbarZ2}
ds^2 = \frac{1}{z^2} \left( -dt^2 + dx^2 + dz^2 \right), \qquad \text{such that }z > z_c, \quad x \geq 0.
\end{align}
Such a metric is equivalent to considering a tensionless EOW brane at $x=0$, as elaborated in \cite{Kastikainen:2021ybu,Aharony:2010ay}. As the boundary of the $Z_2$ quotient $T\bar{T}$ deformed CFT$_2$ moves from $z \to 0$ to some cut-off $z_c$, the EOW brane in the bulk now starts from $z=z_c$. This modifies the transformations in \cref{trans_3} as
\begin{align}\label{trans_4}
z = z_c + y \sech \rho , \qquad x = y \tanh \rho , 
\end{align}
as a result of which the metric in \cref{met_ttbarZ2} becomes
\begin{align}
ds^2 = \frac{1}{\left( z_c + y \sech \rho \right)^2} \left( d \tau_p^2 + dy^2 + \frac{y^2}{\cosh ^2 \rho} d\rho ^2 \right).
\end{align}
Assuming that the EOW brane is located at some $\rho=\rho_0$, the induced metric on the EOW brane may now be expressed as a conformally flat metric with a conformal factor $\Omega_y$ as
\begin{align}\label{eq_cf3}
ds_{brane}^2 = \Omega_y^2 (d \tau_p^2 + dy^2), \qquad \Omega_y= \Big| \frac{1}{z_c + y \sech \rho_0} \Big|,
\end{align}
where we observe that the conformal factor now differs from that in \cref{met_brane} to incorporate the radial cut-off $z_c$ (and in turn the $T\bar{T}$ deformation parameter $\mu$ via \cref{eq_defpar}).

As elaborated earlier in \cref{ssec_defect}, incorporating conformal matter localised on a tensionless EOW brane induces a tension on the brane, which in turn relocates the brane to some constant angle. To impose transparent boundary conditions for the $T\bar{T}$ deformed bath, we include the same field theory on the EOW brane with an identical deformation parameter to that of the $T\bar{T}$ deformed CFT on the cut-off boundary. As discussed in \cite{Deng:2023pjs,Basu:2024xjq} inclusion of this conformal matter to the EOW brane does not backreact on the bulk geometry. This allows us to assume that the cut-off prescription of the holographic dual of the $T\bar{T}$ deformed CFTs is still applicable in the large $N$ approximation.

\subsubsection*{Deformation parameter $\mu$ in the AdS black string coordinates}

From \cref{eq_defpar} it follows that the $T\bar{T}$ deformation parameter is related to the finite cut-off  in the Poincar\'e coordinates as
\begin{align}
\mu \propto z_c^2.
\end{align}
Using \cref{trans_1}, this proportionality may be equivalently expressed in terms of the finite cut-off in the AdS black string coordinates as
\begin{align}\label{eq_murho}
\mu \propto e^{-2\rho_c}.
\end{align}
Note that $z_c$ is a UV regulator, while $\rho_c$ is an IR cut-off. Thus, small values of $\mu$ correspond to small $z_c$ and large $\rho_c$ respectively.

\subsection{Entanglement Entropy}

In this section we compute the first order correction to the entanglement entropy due to $T\bar{T}$ deformation. We once again consider a radiation subsystem $A \equiv [L,u_h] \cup [u_h,R]$ on a constant time slice in the eternal black hole model discussed previously, with the endpoints of the subsystem $A$ now at $P \equiv (t,L,\rho_c)$ and $Q \equiv (-t+i 2 \pi u_h,R,\rho_c)$. Here $\rho_c$ is a very large yet finite, and denotes the new location of the dual field theory as a result of $T\bar{T}$ deformation. For simplicity of computations, we once again begin in the Poincar\'e coordinates, where the corresponding endpoints of $A$ are now at $P \equiv (\tau_1,x_1,z_{c_1})$ and $Q \equiv (\tau_2,x_2,z_{c_2})$.\footnote{Note that is contrast to \cref{ssec_ee}, $z_{c_1}$ and $z_{c_2}$ are now small but finite, and represent the locations of the of the dual field theory due to $T\bar{T}$ deformation.} We later transform the final results using \cref{trans_1,trans_2,trans_3} and express the entanglement entropy in terms of the AdS black string coordinates. Similar to \cref{ssec_ee}, we come across two distinct entanglement entropy phases depending on the distance of the subsystem $A$ from the EOW brane.

\subsubsection{Island Phase}

\begin{figure}[t]
\centering
\includegraphics[scale=0.8]{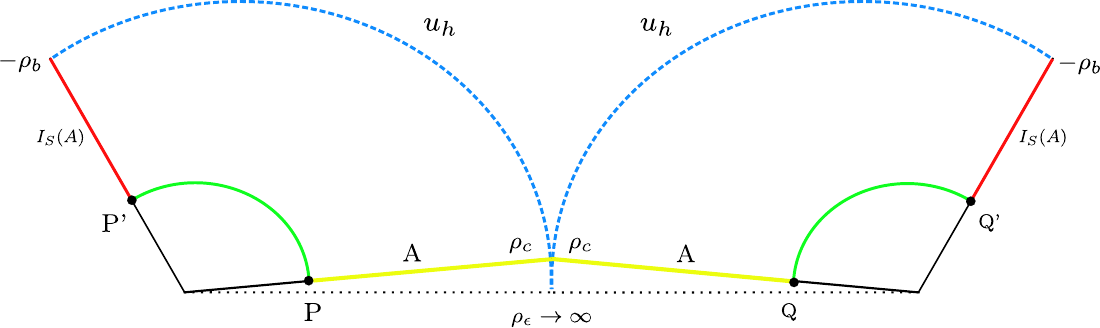}
\caption{The Island phase of the entanglement entropy in the presence of $T\bar{T}$ deformation. The radiation subsystem $A$ is denoted by the yellow subregions on a cut-off boundary at $\rho_c$, while the corresponding entanglement entropy island $I_S(A)$ are illustrated by the red regions on the EOW brane. The green curves represent the RT surfaces.}\label{fig_eti}
\end{figure}

\textbf{Field theory computation :} As discussed in the previous section, this phase corresponds to the scenario where the points $P$ and $Q$ are located close to the brane such that the radiation subsystem $A$ has a corresponding island region on the brane. This is illustrated in \cref{fig_eti}. The entanglement entropy in this case may be computed using \cref{eq_eeis}, where the effective term may be evaluated using the standard form of the 2-point correlator of twist fields at points $P,P',Q$ and $Q'$ as
\begin{align}
S_{eff} & = \lim_{n \to 1} \frac{1}{1-n} \Big[ \log \Omega_{P'}^{-2 h_n}\langle \sigma_n (P) \bar{\sigma}_n (P') \rangle \uc + \log \Omega_{Q'}^{-2 h_n}\langle \sigma_n (Q') \bar{\sigma}_n (Q) \rangle \uc \Big] \notag \\
& = \frac{c}{6} \left( \log \left[ \frac{(y_{_{P'}}+x_1)^2+(\tau_1-\tau_{_{P'}})^2}{z_{c_1} \epsilon_y} \right] + \log \left[ \frac{\ell}{y_{_{P'}} \sech \rho _b+z_{c_1}} \right] \right) \notag \\
& \qquad + \frac{c}{6} \left( \log \left[ \frac{(y_{_{Q'}}+x_1)^2+(\tau_2-\tau_{_{Q'}})^2}{z_{c_2} \epsilon_y} \right] + \log \left[ \frac{\ell}{y_{_{Q'}} \sech \rho _b+z_{c_2}} \right] \right),
\end{align}
where the conformal factors $\Omega_{P'}$ and $\Omega_{Q'}$ are now given by \cref{eq_cf3}, and have once again assumed $P' \equiv (-y_{_{P'}},\tau_{_{P'}})$ and $Q' \equiv (-y_{_{Q'}},\tau_{_{Q'}})$. Due to the presence of the island region on the brane, the contributions due to the area term in \cref{eq_eeis} must also be taken into consideration. This term may be directly obtained from \cite{Deng:2023pjs} as 
\begin{align}
S_{area} & = \frac{1}{4 G_N} \cosh ^{-1}\left[\frac{\sqrt{2 z_{c_1} y_{_{P'}} \sech \rho _b+y_{_{P'}}^2+z_{c_1}^2}}{y_{_{P'}} \sech \rho _b+z_{c_1}}\right] \notag \\
& \qquad \qquad + \frac{1}{4 G_N} \cosh ^{-1}\left[\frac{\sqrt{2 z_{c_2} y_{_{Q'}} \sech \rho _b+y_{_{Q'}}^2+z_{c_2}^2}}{y_{_{Q'}} \sech \rho _b+z_{c_2}}\right] .
\end{align}
The generalized entanglement entropy $S_{gen}$ is then obtained by adding $S_{eff}$ and $S_{area}$ computed above. Extremizing $S_{gen}$ upto first order in $z_{c_1}$ and $z_{c_2}$, we obtain $\tau_{_{P'}}=\tau_1, y_{_{P'}} = x_1 -2 z_{c_1} e^{\rho_b}$ and $\tau_{_{Q'}}=\tau_2, y_{_{Q'}} = x_2 -2 z_{c_2} e^{\rho_b}$, and the final entanglement entropy (upto first order corrections in $z_1$ and $z_2$) may then be obtained as
\begin{align}\label{eq_eeipt}
S(A) & = \frac{c}{6} \left( \log \frac{2 x_1}{z_{c_1}} + \log \frac{2 x_2}{z_{c_2}} + 2 \rho_b + 2 \log  \frac{2 \ell}{\epsilon_y \sech \rho _b} \right)-\frac{c}{6} \frac{z_{c_1} e^{\rho_b}}{x_1} -\frac{c}{6} \frac{z_{c_2} e^{\rho_b}}{x_2} \notag \\
& = \frac{c}{3} \left( \rho_c + \rho_b + \log  \frac{2 \ell}{\epsilon_y \sech \rho _b}  \right) -\frac{2 c}{3} e^{\rho_b-\rho_c},
\end{align}
where in the final step we utilize the transformations in \cref{trans_1,trans_2,trans_3} to express the results in the AdS black string coordinates. Note that in the limit $\rho_c \to \rho_{\epsilon}$, the dominant term (the first term in \cref{eq_eeipt}) is in exact agreement with the results of the undeformed scenario given by \cref{eq_eeip}.

\vspace{0.5cm}

\textbf{Bulk computation :} From the bulk perspective the entanglement entropy may be determined using the DES formula in \cref{eq_eedes}, where the first term may be evaluated in terms of the geodesic lengths of the two RT surfaces $PP'$ and $QQ'$ using \cref{eq_geol} as
\begin{align}
S_{RT} = & \frac{1}{4 G_N} \cosh ^{-1} \left[ \frac{(x_1 + y_{_{P'}} \tanh \rho_b)^2 + (z_{c_1} + y_{_{P'}} \sech \rho_b)^2 + (\tau_1-\tau_{_{P'}})^2 + z_{c_1}^2}{2 z_{c_1} (z_{c_1} + y_{_{P'}} \sech \rho_b)} \right] \notag \\
& \qquad + \frac{1}{4 G_N} \cosh ^{-1} \left[ \frac{(x_2 + y_{_{Q'}} \tanh \rho_b)^2 + (z_{c_2} + y_{_{Q'}} \sech \rho_b)^2 + (\tau_2-\tau_{_{Q'}})^2 + z_{c_2}^2}{2 z_{c_2} (z_{c_2} + y_{_{Q'}} \sech \rho_b)} \right],
\end{align}
where we now utilize \cref{trans_4} to express the coordinates of the brane points as $P' \equiv (\tau_{_{P'}},-y_{_{P'}} \tanh \rho_b, z_{c_1} + y_{_{P'}} \sech \rho_b)$ and $Q' \equiv (\tau_{_{Q'}},-y_{_{Q'}} \tanh \rho_b, z_{c_2} + y_{_{Q'}} \sech \rho_b)$. The defect term in \cref{eq_eedes} contributes in this phase due to the presence of entanglement entropy islands on the brane, which may be directly obtained from \cite{Deng:2023pjs}. The generalized entanglement entropy may then be determined as
\begin{align}
S_{gen} = S_{eff} + \frac{c}{6} \log \frac{2 y_{_{P'}}}{z_{c_1} (z_{c_1} + y_{_{P'}} \sech \rho_b)} + \frac{c}{6} \log \frac{2 y_{_{Q'}}}{z_{c_2} (z_{c_2} + y_{_{Q'}} \sech \rho_b)}.
\end{align}
Extremizing the above expression we obtain, upto first order in $z_{c_1}$ and $z_{c_2}$, $\tau_{_{P'}}=\tau_1, y_{_{P'}} = x_1 -2 z_{c_1} e^{\rho_b}$ and $\tau_{_{Q'}}=\tau_2, y_{_{Q'}} = x_2 -2 z_{c_2} e^{\rho_b}$. The final holographic entanglement entropy may then be obtained as \cref{eq_eeipt} on application of the transformations in \cref{trans_1,trans_2,trans_3} and the Brown-Henneaux relation.

\subsubsection{No-Island Phase}

\begin{figure}[t]
\centering
\includegraphics[scale=0.8]{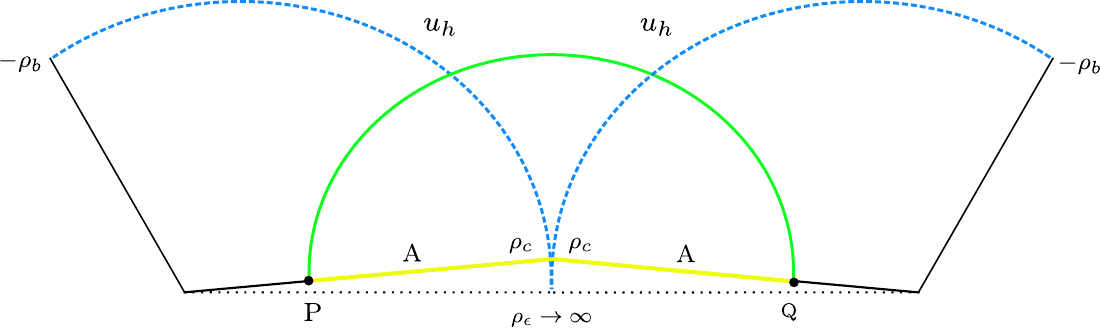}
\caption{The No-Island phase of the entanglement entropy in the presence of $T\bar{T}$ deformation. The radiation subsystem is once again denoted by the yellow subregions on a cut-off boundary at $\rho_c$. The RT surface in this scenario is a Hartman-Maldacena type surface, and is represented by the green curve.}\label{fig_etni}
\end{figure}

\textbf{Field theory computation :} This phase corresponds to the scenario where the points $P$ and $Q$  are located far away from the EOW brane, as a result of which the radiation subsystem $A$ has no corresponding entanglement entropy island on the brane, as illustrated in \cref{fig_etni}. Consequently, the area term in \cref{eq_eeis} does not contribute and the entanglement entropy is equivalent to the effective term in this case, which may be determined in terms of the 2-point correlator of twist fields at $P$ and $Q$ as given in \cref{eq_pqcorr}. The rest of the computation follows as elaborated in \cref{ssec_eeni}, and the final entanglement entropy may then be expressed as 
\begin{align}\label{eq_eenipt}
S(A) = \frac{c}{6} \log \left[ \frac{u_h}{L R} \left( \Delta_L+\Delta_R + 2 \sqrt{\Delta_L \Delta_R} \cosh \frac{t}{u_h} \right) \right]+ \frac{c}{3} \rho_c.
\end{align}
We observe no correction to the entanglement entropy in this phase due to $T\bar{T}$ deformation.

\vspace{0.5cm}

\textbf{Bulk computation :} Utilizing the DES formula in \cref{eq_eedes}, the holographic entanglement entropy in the no-island phase is simply given by the first term. As elaborated earlier, due to the absence of entanglement entropy islands on the EOW brane the defect term does not contribute in this scenario. The holographic entanglement entropy, which in this case is simply $S_{RT}$, is thus obtained as 
\begin{align}
S(A) = S_{RT}=\frac{1}{4 G} \cosh ^{-1} \left[ \frac{(x_2-x_1)^2+(\tau_2-\tau_1)^2+z_{c_1}^2+z_{c_2}^2}{2 z_{c_1} z_{c_2}} \right],
\end{align}
which matches with \cref{eq_eenipt} on application of the transformations described in \cref{trans_1,trans_2,trans_4} and the Brown-Henneaux relation.

\section{Page curve}\label{sec_pc}

\begin{figure}[t]
\centering
\includegraphics[scale=0.5]{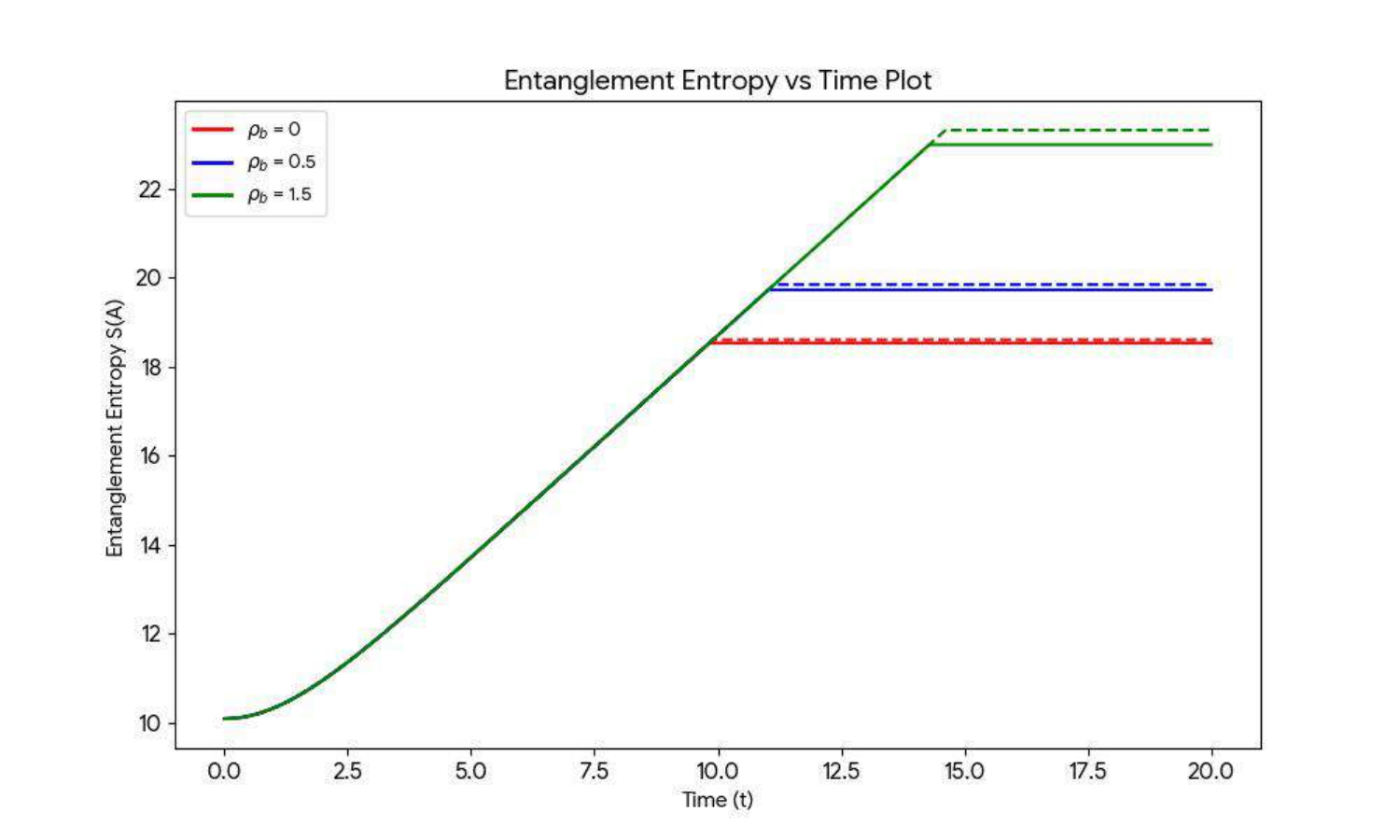}
\caption{This plot represents the Entanglement Entropy Page curve for the radiation subsystem $A$, for both undeformed (depicted by the dashed curves) and $T\bar{T}$ deformed (depicted by the solid curves) scenarios. The Page curves corresponding to distinct values of the brane angle $\rho_b$ are depicted using different colors. We set $c=6,u_h=1, \epsilon_y=0.01,\rho_{\epsilon}=\rho_c=4$, and $L=R=0.5$.}\label{fig_page curve}
\end{figure}

In this section we plot and analyse the Page curves for the entanglement entropy of the radiation subsystem $A$, for both undeformed and $T\bar{T}$ deformed scenarios. To this end, we evaluate the entanglement entropy under the assumption $c=6,u_h=1,\epsilon_y=0.01,L=R=0.5$, and $\rho_{\epsilon}=\rho_c=4$.\footnote{We assume that $\rho_c$ and $\rho_{\epsilon}$ to have same values in this scenario. This is because the objective of this analysis is to investigate the changes in the original Page curve when corrections due to $T\bar{T}$ deformation is taken into account. For this purpose we require the undeformed entanglement entropy to have the same value numerically for both scenarios.} In addition we systematically vary $\rho_b$ to observe how the Page curve changes with the brane angle. This is illustrated in \cref{fig_page curve}, where the Page curves corresponding to distinct brane angles are depicted using different colors. The Page curve for the deformed scenarios are depicted by the solid curves, while those for the undeformed scenarios are depicted by the dashed curves. We observe that for both scenarios the entanglement entropy increases with time in the no-island phase. Upon transition to the island phase the entanglement entropy stabilizes to a constant value, which varies with change in the brane angle.

A general trend emerges from \cref{fig_page curve}. We observe in the no-island phase the Page curves overlap for both undeformed and $T\bar{T}$ deformed scenarios and for all values of $\rho_b$, indicating that the evolution of the entanglement entropy is independent of the deformation parameter or the brane angle in this phase. However, we observe some interesting features in the island phase. The entanglement entropy in this phase is dependent on both the brane angle and the deformation parameter, with the deformed entanglement entropy being smaller than the undeformed one for a given brane angle. Moreover, the Page time $T_p$ (the time of transition between the no-island and the island phases of the entanglement entropy) appears to be dependent on the brane angle as well as the deformation parameter. This dependence can be mathematically explored by deriving the expressions of $T_p$ for both the undeformed and the $T\bar{T}$ deformed scenarios as
\begin{align}\label{eq_pt}
T_p = 
\begin{cases}
u_h \cosh^{-1} \left[ \frac{L R \left(e^{2 \rho_b}+1\right)^2 -u_h \epsilon_y^2 (\Delta_L + \Delta_R)}{2 u_h \epsilon_y^2 \sqrt{\Delta_L \Delta_R}} \right] \qquad
& \text{Undeformed scenario,} \\
u_h \cosh^{-1} \left[ \frac{L R \left(e^{2 \rho_b}+1\right)^2 e^{-4 e^{\rho_b-\rho_c}} -u_h \epsilon_y^2 (\Delta_L + \Delta_R)}{2 u_h \epsilon_y^2 \sqrt{\Delta_L \Delta_R}} \right] \qquad
& T\bar{T}\text{-deformed scenario,}
\end{cases}
\end{align}
where the Page time for the undeformed scenario may be directly obtained from that of the deformed scenario by taking the limit $\rho_c \to \infty$.

\begin{figure}[t]
\centering
\includegraphics[scale=0.33]{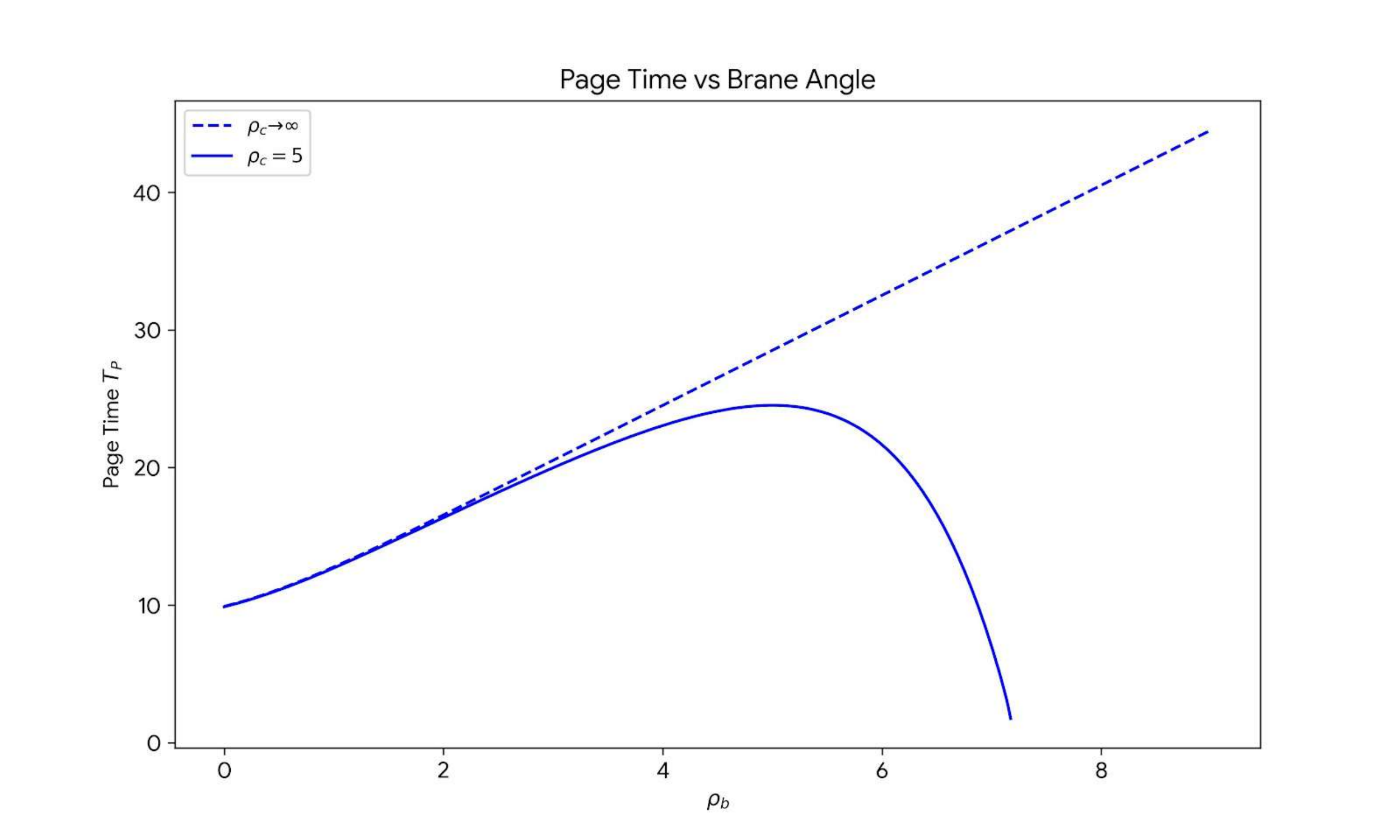}
\caption{This plot represents how the Page time $T_p$ changes with the brane angle. The dashed curve represents the undeformed scenario, while the solid curve represents the $T\bar{T}$ deformed scenario. We have set $u_h=1, \epsilon_y=0.01$, and $L=R=0.5$.}\label{fig_Tpbrane}
\end{figure}

It is interesting to further investigate the dependence of the Page time $T_p$ on the brane angle and the deformation parameter and explore any new physics that might emerge. We explore the dependence of $T_p$ on the brane angle in \cref{fig_Tpbrane} for both the undeformed and the $T\bar{T}$ deformed scenarios. Note that though the Page time initially increases with $\rho_b$, it rapidly drops to zero for the $T\bar{T}$ deformed scenarios as $\rho_b$ further increases. This is in contrast with the undeformed scenario, where the Page time increases consistently with $\rho_b$. This behaviour is also observed in \cite{Basu:2024xjq}, where the authors speculate that this rapid decrease in the Page time might be due the CFT being no longer located at the asymptotic boundary at $\rho \to \infty$, but pushed inwards to a finite value in the presence of $T\bar{T}$ deformation.

Also note that for deformed scenarios, there seems to be an upper bound to the physically admissible values of $\rho_b$ for which $T_p$ is strictly positive, in contrast to the undeformed scenario where the brane angle may be increased without bound. Interestingly, this upper bound may itself exceed the finite cut-off arising from $T\bar{T}$ deformation, a feature which requires further investigation.

\begin{figure}[t]
\centering
\includegraphics[scale=0.33]{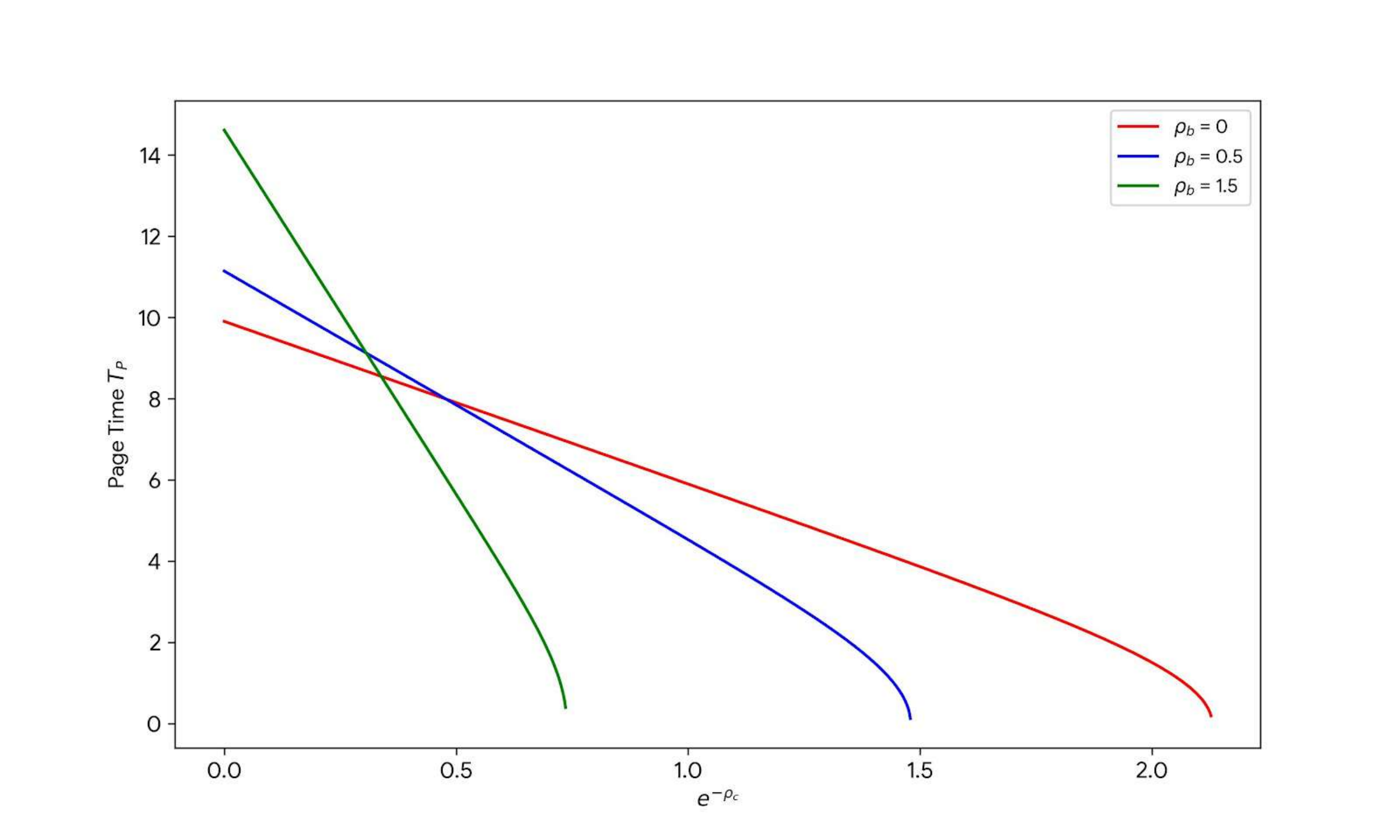}
\caption{This plot represents how the Page time $T_p$ changes with the deformation parameter for different brane angles $\rho_b$. We have set $u_h=1, \epsilon_y=0.01$, and $L=R=0.5$.}\label{fig_Tpcutoff}
\end{figure}

We further extend the analysis by examining how the Page time changes with the deformation parameter in \cref{fig_Tpcutoff}, where we plot $T_p$ against $e^{-\rho_c}$ (related to the deformation parameter as given in \cref{eq_murho}) for different brane angles. $e^{-\rho_c} \to 0$ corresponds to the undeformed scenario, while increasing values indicate $T\bar{T}$ deformed scenarios. We observe that for all brane angles the Page time decreases approximately linearly with the UV cut-off $e^{-\rho_c}$, but shows a sudden drop at the end. With increase in $\rho_b$, $T_p$ starts at a higher value, but decreases much rapidly for high brane angles compared to low brane angles.

\section{Summary and Discussions}\label{sec_summary}

In this article, we investigate the proposed duality between the island and the defect extremal surface (DES) prescription for the fine-grained entanglement entropy in a Karch-Randall (KR) braneworld model, where the radiation baths are located in a gravitational background. In particular, we consider an AdS$_3$ black string geometry truncated by an end-of-the-world (EOW) brane for which the effective lower-dimensional perspective consists of a radiation bath defined on an AdS$_2$ black hole background. We consider subsystems on radiation baths corresponding to AdS$_2$ eternal black holes in the lower-dimensional effective picture, and compute the entanglement entropy for distinct phases using the island formula in the effective $2d$ and the DES formula in the bulk picture. We demonstrate an agreement between the two prescriptions, thus establishing that the DES formula is a holographic dual to the island formula. Additionally, the results agree with those existing in the literature obtained using the bulk and the boundary prescription \cite{Geng:2021mic,Geng:2022dua}, which acts as a check for the accuracy of our computations.

We further investigate the entanglement structure for AdS$_3$ black string geometry with a finite cut-off arising from $T\bar{T}$ deformation \cite{McGough:2016lol}, truncated by an EOW brane. The usual AdS/CFT duality applies to the Dirichlet boundary on the cut-off surface, while Neumann boundary conditions apply on the EOW brane \cite{Deng:2023pjs}. We consider an eternal AdS$_2$ black hole coupled to radiation baths (now located on a finite cut-off surface induced by $T\bar{T}$ deformation), and determine the entanglement entropy up to first order for radiation subsystems using both the island and the DES formulas. We observe an agreement between the two prescriptions for both the island and no-island phases of the entanglement entropy, providing another strong consistency check for the proposed duality. 

Subsequently, we construct and compare the entanglement entropy Page curves for both undeformed and deformed scenarios discussed above. While the entanglement entropy for the no-island phase remains unchanged up to first order in $e^{-\rho_c}$ (related to the deformation parameter) in the presence of $T\bar{T}$ deformation, a significant correction depending on the angle of the brane $\rho_b$ is observed for the island phase of the entanglement entropy. Moreover, the Page time $T_p$ (time of transition between the no-island to the island phases of the entanglement entropy), is also dependent on $\rho_b$ and appears to be modified by the $T\bar{T}$ deformation. We further examine the dependence of $T_p$ on $\rho_b$ and the finite cut-off $\rho_c$ and uncover some interesting features. Note that the results and plots provided in this article are valid strictly in the perturbative limit, and may differ when higher order corrections are considered. 

Though \cite{Chen:2018eqk,Murdia:2019fax,He:2019vzf} have established that modifying the correlator by appropriate insertions of stress tensor components correctly captures the effects of the $T\bar{T}$ deformation, in this article we adopt a different method for our analysis. Similar to \cite{Deng:2023pjs}, we assume that this deformation manifests through modifications in the induced metric defined on the radial cut-off surface and observe an agreement between the results obtained using the island and the DES prescriptions. However, a perturbative theory of $T\bar{T}$ deformed BCFT must still be developed.

This work identifies several promising directions to explore further. An important consideration is analysing the mixed state entanglement structure for both undeformed and deformed scenarios for KR braneworld models. Extending the framework to higher dimensions is an equally crucial advancement, particularly in light of \cite{Chu:2021gdb}, which shows that the DES and the island prescriptions do not align consistently for higher dimensions even for undeformed scenarios. Furthermore, testing the proposed duality between these prescriptions for other deformed theories is also a compelling direction. We leave these open questions for future study.

\section*{Acknowledgement}

I would like to thank my colleagues Saikat Biswas, Ankit Anand, and Himanshu Chourasiya for their valuable feedback.

%%%%%%%%%%%%%%%%%
%%%%%%%%%%%%%%%%%
%%%%%%%%%%%%%%
%%%%%%%%%%%%%%%%%%%

\bibliographystyle{utphys}

\bibliography{ref}

\end{document}